\documentstyle[prl,aps,psfig]{revtex}

\draft
\def\noi{\noindent}
\def\cm{\,{\rm cm}}
\def\km{\,{\rm km}}

\def\mpc{\,{\rm Mpc}}

\def\ev{\,{\rm eV}}

\def\gev{\,{\rm GeV}}
\def\tev{\,{\rm TeV}}

\def\km{\,{\rm km}}
\def\sec{\,{\rm sec}}

\def\la{\mathrel{\mathpalette\fun <}}
\def\ga{\mathrel{\mathpalette\fun >}}
\def\fun#1#2{\lower3.6pt\vbox{\baselineskip0pt\lineskip.9pt
  \ialign{$\mathsurround=0pt#1\hfil##\hfil$\crcr#2\crcr\sim\crcr}}}
\newcommand{\beq}{\begin{equation}}
\newcommand{\eeq}{\end{equation}}
\begin{document}
\twocolumn[\hsize\textwidth\columnwidth\hsize\csname
@twocolumnfalse\endcsname


\draft
{\vskip-2.5truecm{\flushright{\hfill Bartol Preprint BA-97-48,
hep-ph/9710533 (To appear in Phys. Rev. Lett.)}}} 
\title{TeV and Superheavy Mass-Scale Particles
from Supersymmetric Topological Defects, the Extragalactic 
$\gamma$-ray Background, and the Highest Energy Cosmic Rays}
   
\author{Pijushpani Bhattacharjee$^{1,2}$, Qaisar Shafi$^3$, 
and F. W. Stecker$^1$}

\address{
$^1$Laboratory for High Energy Astrophysics, Code 661,
NASA/Goddard Space Flight Center, Greenbelt, MD 20771, USA.}

\vspace{0.8cm}

\address{$^2$Indian Institute of Astrophysics, Bangalore - 560 034,
INDIA.}

\vspace{0.8cm}

\address{
$^3$Bartol Research Institute, University of Delaware, Newark,
Delaware 19716, USA.}

\maketitle

\begin{abstract}

Cosmic topological defects in a wide class of  
supersymmetric theories can simultaneously be sources of higgs
particles of mass comparable to the supersymmetry breaking scale
$\sim\tev$, as well as superheavy gauge bosons of mass $\sim\eta$, 
where $\eta$ ($\gg 1\tev$) denotes the associated gauge symmetry breaking 
scale. For cosmic strings with $\eta\sim10^{14}\gev$, the higgs decay 
can account for the extragalactic diffuse $\gamma$-ray
background above $\sim$ 10 GeV, while the gauge boson decay may explain 
the highest energy cosmic ray flux above $\sim10^{11}\gev$, {\it
provided that} particle 
production is the dominant energy loss mechanism for cosmic strings, as
recent simulations suggest. By the same token, cosmic strings with 
$\eta$ much above $\sim10^{14}\gev$ will be ruled out. 

\end{abstract}

\vskip 2pc]

\noi 
In a wide class of supersymmetric (SUSY) unified gauge theories, 
including some versions of effective theories derived from
superstrings, certain phase transitions can occur      
at a temperature comparable to the ``soft'' supersymmetry breaking scale
$\la 1\tev$, even though the associated gauge symmetry breaking scale 
itself may be much larger~\cite{shafi1,shafi_mono}. This occurs rather
generically in SUSY theories as a consequence of 
existence~\cite{flat1} of `directions' 
along which the effective potential $V$ of the relevant scalar field is  
almost flat, i.e., the curvature $\vert V^{\prime\prime}\vert^{1/2}$ 
of the potential is much smaller than the 
vacuum expectation value (VEV) $\eta$ of the scalar field 
out to field values $\sim \eta$. 
An ``almost flat potential'' for a (complex) higgs
field $\Phi$ after supersymmetry breaking generally has the
form~\cite{shafi1,shafi_mono,lyth,barreiro}    
$V=V_0 - m_s^2 \phi^2 + \sum_{n=1}^\infty \lambda_n 
m_{\rm Pl}^{-2n}\phi^{2n+4}$, where $\phi\equiv\vert\Phi\vert$ and 
$m_{\rm Pl}\equiv(8\pi G)^{1/2}\approx 2.4\times10^{18}\gev$. 
The $\phi^2$ term arises from
`soft' supersymmetry breaking, so the mass scale $m_s$ is typically 
$\la 1\tev$. The higher order (nonrenormalizable) terms would arise from 
`integrating out' particles of Planck mass scales in a `higher' 
theory such as superstring theory. The flatness of the potential 
is due to absence of the $\lambda\phi^4$ term familiar in non-SUSY
theories. 
Depending on the strengths of the couplings $\lambda_n$, 
the minimum of the potential, i.e., the VEV $\eta$, can 
lie anywhere in the range $\sim 10^9\gev$ to $\sim M_{\rm
GUT}\sim10^{16}\gev$, the grand unification (GUT) scale~\cite{gutscale}. 
The `height' of the potential is $V_0\sim\eta^2 m_s^2$. 

For temperatures
$T$ in the range $m_s\ll T\la V_0^{1/4}$, finite temperature corrections  
to the potential can hold the $\Phi$ field at $\phi=0$ 
until $T$ falls below $m_s$, at which a
phase transition occurs taking $\phi$ to $\eta$. 
Note that although the higgs scalars in these theories are `light' with
mass 
$m_\phi\sim\vert V^{\prime\prime}\vert^{1/2}\sim m_s\la 1\tev$, the
associated gauge bosons have the `usual' mass $\sim\eta$, which can, in
particular, be $\sim M_{\rm GUT}$, if the higgs under consideration breaks  
the GUT symmetry. Some cosmological consequences of theories with almost 
flat potentials (hereafter simply referred to as ``flat SUSY'' theories)
have been considered in Refs.~\cite{shafi_mono,lyth,barreiro}. 

In this Letter, we point out that cosmic
topological defects (TDs)~\cite{tdreview} such as
magnetic monopoles and cosmic strings associated with 
phase transitions in flat SUSY theories~\cite{shafi_mono,barreiro} 
can, through their collapse, annihilation, or other processes, be 
sources of higgs bosons of mass $\sim\tev$, {\it as well as} of
gauge bosons of superheavy mass scale $\sim\eta\gg 1\tev$, 
and that the decay products of both these kinds of particles may be
observable in the Universe today. 

Production of extremely energetic photons, nucleons and
neutrinos through decay of massive ``X'' particles (typically of GUT-scale
mass $\sim10^{16}\gev$) originating from  
TDs~\cite{bhs,br,vincent,mono_necklace,other_tds},    
is a subject of much current interest as a possible 
explanation~\cite{tdhecr,chicago,ps} of 
the highest energy cosmic ray (HECR) events at energies $\ga10^{11}\gev$
\cite{hecr}. These TDs have usually been considered 
within the context of the 
standard non-SUSY (and non-flat) quartic GUT symmetry-breaking higgs
potential~\cite{tdreview}, 
for which the relevant phase transition occurs at a temperature
$T\sim\eta\sim10^{16}\gev$. In this case, the 
associated GUT gauge bosons as well as higgs bosons, and consequently the X
particles `constituting' the TDs, all have masses 
of order the GUT scale VEV $\eta\sim10^{16}\gev$. 
In contrast, the new feature in flat SUSY theories is
that the X particles produced by the same 
TD processes~\cite{bhs,br,vincent,mono_necklace,other_tds} would now be 
higgs of mass $\sim\tev$ as well as superheavy gauge bosons of mass
$\sim\eta\gg1\tev$. We show that decay of these TD-produced TeV mass-scale
higgs may contribute significantly to the extragalactic diffuse
$\gamma$-ray background (EDGRB)~\cite{sreek} above a few GeV
(which seems to be difficult to explain otherwise
in terms of emissions from astrophysical objects; see below), while
the superheavy gauge bosons could be a source of the HECR particles.  
In particular, we show that cosmic strings in flat SUSY theories with
$\eta\sim10^{14}\gev$ may simultaneously explain 
{\it both} EDGRB above a few GeV and HECR,  
if X particle production (rather than gravitational
radiation emission) is the dominant energy loss mechanism for 
cosmic strings --- a possibility recently suggested in
Ref.~\cite{vincent}. By the same token, cosmic strings 
with $\eta$ much larger than $10^{14}\gev$ (and hence  
GUT-scale cosmic strings with $\eta\sim10^{16}\gev$) in flat SUSY theories 
overproduce {\it both} EDGRB and HECR, and are, therefore, ruled out. 
(In this case, cosmic strings with $\eta$ much larger than $10^{14}\gev$
in non-SUSY theories are also ruled out because they overproduce HECR). 

Note that although the higgs in flat SUSY theories are
`light', they are not expected to be produced in accelerators operating at
energies well below the energy scale $\sim\eta$ because their coupling
to minimal supersymmetric standard model (MSSM) particles, for
example, is expected to be suppressed by a factor   
$\sim m_\phi/\eta\ll 1$. Thus, TDs may indeed be the only source
of these higgs in the {\it present day} Universe. 

The rate of X particle production per unit volume at a time $t$ in the
matter dominated epoch from a system of TDs can be generally written in
the form~\cite{bhs} 
$dn_X/dt=(Q_0/m_X)\, (t/t_0)^{-4+p}\,,$
where $t_0$ denotes the present epoch, and $Q_0$ is the rate
of total energy released per unit volume in the form of X particles
(higgs plus gauge bosons) in the present epoch. (We use natural units,
$\hbar=c=1$, throughout.) On general grounds we expect that the total
energy released will be roughly equipartitioned between the higgs
($Q_{0,\phi}$) and the gauge boson ($Q_{0,V}$) modes, and so we will
assume that $Q_{0,\phi}\approx Q_{0,V}\approx (1/2)Q_0$. The dimensionless
parameter $p$ is in general different for different systems of
TDs~\cite{bhs}. Here we consider the case 
$p=1$, which is representative of a large class of
TD processes including those involving cosmic
strings and magnetic monopoles~\cite{br,vincent,mono_necklace}. 
The case in which the X
particles are of heavy mass scales $\sim O(\eta)$ has been considered 
earlier~\cite{bhs,br,mono_necklace,other_tds,tdhecr,chicago,ps}. Here we
consider the effects of the TeV mass-scale higgs X particles, which we
shall assume to be non-relativistic.  
These higgs would decay on a time scale~\cite{shafi1,lyth,barreiro} 
$\tau\sim 6.6\xi^{-1}(\eta/10^{16}\gev)^2(1\tev/m_\phi)^3\sec$, where 
$\xi\la 1$ is a numerical factor~\cite{barreiro}. For relevant values of
$\eta$ and $m_\phi$ this decay is essentially `instantaneous' on 
cosmological time scales at late epochs of interest to us.    

By far the largest number of particles eventually produced by an 
X would come through the hadronic jet fragmentation of quarks and 
gluons resulting from its decay (see, e.g., Ref.~\cite{kribs} for
arguments concerning the dominance of the hadronic decay
channel)~\cite{fn1}. The fragmentation of the quarks/gluons into
jets of hadrons and the photon spectrum resulting from decay of neutral
pions in these jets are well described by the string fragmentation scheme
implemented in the JETSET program~\cite{jetset}. 
We assume typical hadronic 3-body decays~\cite{kribs} of a X particle
into all kinematically available quark pairs and one uncolored (assumed
massless) spectator, and obtain the injected photon spectrum  
from the decay of a single higgs X particle by using the parametrization
of the photon spectrum derived from JETSET, as described in
Ref.~\cite{kribs}. Folding this spectrum with the X particle production
rate then gives us the full injection spectrum. 

In our calculation of the predicted total $\gamma$-ray flux today, 
we have included the effects of electromagnetic cascading and
$\gamma$-$\gamma$ scattering~\cite{cascade,kribs}, and also included 
the effect of absorption due to pair production on
photons of infrared, optical and ultraviolet backgrounds~\cite{irouv}.  
The details of these calculations will be given elsewhere. In our
numerical calculations we have assumed a spatially flat universe with
$\Omega_0=1$ and present Hubble constant $H_0=75 \km \sec^{-1} \mpc^{-1}$. 
\begin{figure} 
\centerline{\psfig{file=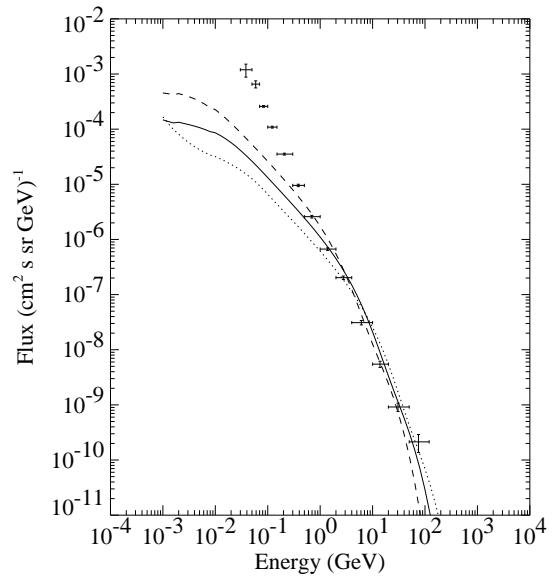,height=3.2in}}
\medskip
\caption[]{Gamma ray flux due to decay of the higgs from TD processes with
$p=1$, 
higgs mass $m_\phi=$ 500 GeV (dashed curve), 1 TeV (solid curve), and
2 TeV (dotted curve), and $Q_{0,\phi}=4.5\times10^{-23}\ev \cm^{-3}
\sec^{-1}$. The extragalactic diffuse $\gamma$-ray
background data from EGRET (Ref.~\cite{sreek}) are also shown for
comparison.}
\label{Fig_1}
\end{figure}  

Fig.1 shows the higgs contribution to the EDGRB for TD processes with
$p=1$ and higgs mass $m_\phi=$ 500 GeV, 1
TeV, and 2 TeV. The normalization of the curves in Fig. 1 corresponds to
$Q_{0,\phi}\simeq 4.5\times10^{-23}\ev \cm^{-3} \sec^{-1}$, 
which, as evident from Fig. 1, is
an upper limit on $Q_{0,\phi}$ (and hence on $Q_0$) for $p=1$ TD
processes imposed by the EGRET data. 

A recent analysis~\cite{sreek} of the EDGRB indicates that the spectrum
continues
at least up to $\sim$ 100 GeV. The EDGRB up to $\sim$ 10 GeV can be
interpreted as arising from a superposition of unresolved
blazars~\cite{stecker_salamon_96}. However, if blazar
$\gamma$-rays are produced by Compton upscattering of lower energy blazar
photons, then only X-ray selected BL Lac objects, a small fraction of the
blazar population, may qualify as possible contributors to EDGRB above 10
GeV, because only they may have the requisite luminosities. 
The majority of radio selected BL Lac objects and flat spectrum radio
quasars are likely to have luminosities falling off above
$\sim$ 10 GeV~\cite{sds_96}. 
Also, extragalactic absorption effects are likely to steepen the
high energy spectra of high-redshift quasars above $\sim$ 10 GeV
\cite{irouv}. Thus, there may be cause to
consider another component of cosmic diffuse $\gamma$-ray
emission. From the shapes of the curves in Fig.1, we see that 
decays of higgs of mass $\sim$ 500 -- 1000 GeV from p=1 TD
processes in flat SUSY theories may  
play an important role in producing the EDGRB in the 10 -- 100 GeV energy
range.  
The rate of energy injection in the form of TeV higgs
needed to explain the EDGRB above a few GeV is $Q_0^{\rm EDGRB}
\simeq 4.5\times10^{-23}\ev \cm^{-3}\sec^{-1}$ for $m_\phi\simeq1\tev$.    

The upper limit on $Q_{0,\phi}$ from EDGRB also implies (through energy
equipartition arguments) an upper limit on $Q_0$ as well as on $Q_{0,V}$,
i.e., $Q_0/2\approx Q_{0,\phi}\approx Q_{0,V}\la
4.5\times10^{-23}\ev\cm^{-3}\sec^{-1}$. There is, however, an
independent upper limit imposed on $Q_{0,V}$ (and hence on
$Q_0$) by the HECR data: $Q_{0,V}\la Q_0^{\rm HECR}
\simeq 3.3\times10^{-22}(\eta/10^{16}\gev)^{0.5}\ev
\cm^{-3}\sec^{-1}$~\cite{fn2}, where $Q_0^{\rm
HECR}$ is the rate of energy injection needed to explain the HECR. 
For $\eta > 1.9\times10^{14}\gev$, we
overproduce EDGRB if we wish to explain the HECR, and is, therefore,
unfavored. (Of course, TDs with $\eta > 1.9\times10^{14}\gev$ can 
give significant contribution to
the EDGRB while not contributing significantly to the
HECR.) For $3\times10^{11}\gev\ll\eta < 1.9\times10^{14}\gev$ 
we can produce the HECR, but in this case we significantly underproduce
EDGRB if $\eta\ll 1.9\times10^{14}\gev$. 
The two independent upper limits can be
saturated, i.e., we can explain both EDGRB above a few GeV and HECR, if
$\eta\approx1.9\times10^{14}\gev$. 

The value of $Q_0$ for a general TD process is not known {\it a priori}
--- it depends on at
least two (not necessarily mutually independent) unknown parameters,
namely, the fraction of the total energy density of the
relevant defects going into X particles, and the symmetry breaking scale
$\eta$ at which the relevant TDs are formed. 
Therefore, the above arguments do {\it not} by themselves rule
out the existence of GUT scale (i.e., $\eta\sim10^{16}\gev$) TDs
in general --- they only tell us that GUT scale TDs are unlikely to be
responsible for HECR, because that would conflict with EDGRB.    

The situation is, however, very different in the case of cosmic strings
with the recent results~\cite{vincent} of numerical simulations of
cosmic string evolution in the Universe. These studies
show that the energy
density, $\rho_s(t)$, in ``long'' (i.e., horizon crossing) strings at any
time $t$ is maintained in the scaling solution~\cite{tdreview}, 
$\rho_s(t)=\mu/(x^2 t^2)$, by energy loss from long strings
occurring predominantly on the scale of the {\it string width}, i.e.
through formation of string width-size small loops which quickly decay
into X particles or through 
direct emission of the X particles that `constitute' the strings, and not
through formation of (sub)horizon size 
loops and their subsequent decay by emission of gravitational
radiation as thought earlier~\cite{tdreview}. (Here $\mu$ is the energy
per unit length of the string, and $x$ is in the range 0.27 --
0.34~\cite{vincent}.) This result, while subject
to confirmation by independent simulations, obviously has important
implications for HECR and EDGRB. Indeed, in  
this case, there is effectively only one free parameter (namely, $\mu$ or
equivalently, $\eta$) in the problem, which also fixes $Q_0$.  
In fact, in this case, the observed data
on ultrahigh energy cosmic rays already rule out~\cite{br,vincent} 
GUT scale cosmic strings for the standard
non-flat potential case. We shall see that this is also true for the flat
potential case, but here an additional constraint (due to 
TeV mass scale higgs) comes from the EDGRB data. 

From the results of Ref.~\cite{vincent},  
the rate of energy loss of strings per unit volume through X particle
emission is $d\rho_X/dt\simeq (2/3)\mu/(x^2 t^3)$, which gives $Q_0\simeq
7.4\mu/t_0^3$, with $x\simeq 0.3$. 
Requiring $Q_{0,\phi}\approx 0.5 Q_0\la Q_0^{\rm
EDGRB}\simeq4.5\times10^{-23}\ev\cm^{-3} \sec^{-1}$, we get 
$\mu\la 4.5\times10^{-5}\times(10^{16}\gev)^2$. 
Taking, for flat potentials, $\mu\sim0.1\eta^2$~\cite{barreiro},
we get $\eta\la 2.1\times10^{14}\gev$. Thus, in this case, GUT scale
cosmic strings necessarily overproduce EDGRB, and are, therefore, ruled
out. A similar constraint follows from HECR:
Here one requires 
$Q_{0,V}\approx 0.5 Q_0\la Q_0^{\rm HECR}\simeq
3.3\times10^{-22}(\eta/10^{16}\gev)^{0.5}\ev \cm^{-3} \sec^{-1}$, 
which gives $\eta\la2.2\times10^{14}\gev$. 
 
Note that for a non-flat potential, where $\mu\simeq\pi\eta^2$
and where one expects the X particles to be predominantly of heavy mass
scale $\sim\eta$, so that 
$Q_{0,V}\approx Q_0$, the constraint on $\eta$
from HECR is $\eta\la1.4\times10^{13}\gev$.  

It is thus clear that cosmic strings with $\eta$ much greater than
$10^{14}\gev$, and in particular, GUT scale cosmic strings with
$\eta\sim10^{16}\gev$, are ruled out both for flat as well as non-flat
potentials {\it if} X particle production is their dominant energy loss
mechanism. At the same time, in this case, 
cosmic strings with $\eta\sim2\times10^{14}\gev$ in SUSY models with flat
potentials  
can potentially account for the high energy ends of both EDGRB
and HECR. In this respect, absence of free parameters other than the
symmetry breaking scale $\eta$ seems to make cosmic strings a 
``natural'' candidate source of HECR (and possibly of EDGRB above a few
GeV). 

Cosmic string formation at
$\eta\sim10^{14}\gev$ rather than at the GUT scale of $\sim10^{16}\gev$ is
not hard to implement. For example, in a SUSY theory,  
the breaking ${\rm SO}(10)\to{\rm SU}(3)\times
{\rm SU}(2)\times{\rm U}(1)_{\rm Y}\times{\rm U}(1)$ can take place at the
GUT unification scale $M_{\rm GUT}\sim10^{16}\gev$ without any cosmic
string formation, but the second U(1)
can be subsequently broken with a flat potential with a VEV 
$\eta\sim10^{14}\gev$ to yield cosmic strings
that are relevant for EDGRB and HECR.   

Cosmic strings with $\eta\sim10^{14}\gev$ would be too light to 
be relevant for structure formation in the Universe. 
Their signatures on the CMBR sky would also be too weak to detect.
However, signatures of these cosmic strings may be searched for with 
next generation $\gamma$-ray instruments such as
GLAST~\cite{glast}, which will be able to resolve further the 
discrete source component and thereby reveal the possible existence of a
truly cosmic component of the EDGRB as in Fig. 1,  
and in proposed HECR observatories such as Auger~\cite{auger} 
and OWL~\cite{owl}. 

We wish to thank P. Sreekumar for stimulating discussions on
EDGRB, and Graham Kribs, Subir Sarkar and G\"unter Sigl for useful
correspondence.  
P.B. acknowledges support under a NAS/NRC Senior Research Associateship
at NASA/GSFC. Q.S. acknowledges the DOE support under grant DE-FG02-91ER40626.


\begin{references}
\bibitem{fn1} Some fraction of the X particle decay products could also be
LSPs, the lightest SUSY particles; but they are not relevant here for our
consideration of the EDGRB. We also neglect other possible non-hadronic
decay channels of X. These approximations may lead to uncertainties
in our flux estimates by factors of order unity, which we think is not 
crucial to our main conclusions.
\bibitem{fn2} This value corresponds to an injection spectrum 
$\propto E^{-1.5}$ and corresponds roughly to the most
conservative bound obtained by
PS~\cite{ps}. One should keep in mind that the value of $Q_0^{\rm HECR}$
is uncertain due to a variety
of uncertain factors including strength of intergalactic magnetic
field, the spectrum of the universal radio background, the injection
spectrum and so on, all of which affect the theoretically calculated
flux. It also depends somewhat on the way one normalizes 
the theoretically predicted flux with the data; see, e.g., 
SLSC~\cite{chicago}.
\end{references}
\end{document}